 \documentclass[final,5p,times,twocolumn,authoryear]{elsarticle}

\usepackage{amssymb}
\usepackage{lipsum}

\usepackage{amsmath}
\usepackage{makecell}
\usepackage{subcaption}
\usepackage{siunitx}
\usepackage{ulem}

\usepackage{algorithm}
\usepackage{algpseudocode}
\usepackage{amsmath}
\algnewcommand\LongComment[1]{%
    \Statex \(\triangleright\) #1
}

\usepackage{hyperref}

\usepackage{xcolor}

\usepackage{tabularx}
\usepackage{booktabs,siunitx,threeparttable,multirow}
\usepackage{fontawesome5}

\usepackage{pgfplots} 
\usetikzlibrary{external}
\makeatletter
\newcommand{\github}[1]{%
   \href{#1}{\faGithubSquare}%
}
\makeatother

\journal{Astronomy $\&$ Computing}

\begin{document}

\begin{frontmatter}



\title{\texttt{gCAMB}: A GPU-accelerated Boltzmann solver for next-generation cosmological surveys}


\author[unich,infnpg]{Loriano Storchi\corref{cor1}}
\ead{loriano@storchi.org, loriano.storchi@unich.it, loriano.storchi@pg.infn.it}
\author[infnferrara,icsc]{Paolo Campeti\corref{cor1}}
\ead{paolo.campeti@fe.infn.it}
\author[infnferrara]{Massimiliano Lattanzi}
\author[unich]{Nicol\'o Antonini}
\author[infnferrara]{Enrico Calore}
\author[infnpg]{Pasquale Lubrano}

\affiliation[infnpg]{organization={INFN sezione di Perugia},
            addressline={Via Pascoli},
            city={Perugia},
            postcode={06123},
            country={Italy}}
\affiliation[unich]{organization={Dipartimento di Farmacia, Università degli Studi G. D’Annunzio},
            addressline={Via dei Vestini},
            city={Chieti},
            postcode={66100},
            country={Italy}}
\affiliation[infnferrara]{organization={INFN Sezione di Ferrara},
            addressline={Via Saragat 1}, 
            city={Ferrara},
            postcode={44122}, 
            country={Italy}}
\affiliation[icsc]{organization={ICSC Centro Nazionale “High Performance Computing, Big Data and Quantum Computing”},
            addressline={Via Magnanelli},
            city={Casalecchio di Reno},
            postcode={40033},
            country={Italy}}
\cortext[cor1]{Author to whom correspondence should be addressed}

\begin{abstract}
Inferring cosmological parameters from Cosmic Microwave Background (CMB) data requires repeated and computationally expensive calculations of theoretical angular power spectra using Boltzmann solvers like \texttt{CAMB}. This creates a significant bottleneck, particularly for non-standard cosmological models and the high-accuracy demands of future surveys. While emulators based on deep neural networks can accelerate this process by several orders of magnitude, they first require large, pre-computed training datasets, which are costly to generate and model-specific. To address this challenge, we introduce \texttt{gCAMB}, a version of the \texttt{CAMB} code ported to GPUs, which preserves all the features of the original CPU-only code. By offloading the most computationally intensive modules to the GPU, \texttt{gCAMB} significantly accelerates the generation of power spectra, saving massive computational time, halving the power consumption in high-accuracy settings and, among other purposes, facilitating the creation of extensive training sets needed for robust cosmological analyses. We make the \href{https://github.com/lstorchi/CAMB/tree/gpuport}{\texttt{gCAMB} \faGithub} software available to the community. 
\end{abstract}



\begin{keyword}
Numerical cosmology \sep Cosmic microwave background \sep GPU acceleration \sep High performance computing  \sep Heterogeneous computing \sep Parallel algorithms \sep Energy-efficient computing 


\end{keyword}

\end{frontmatter}




\section{Introduction}
\label{sec:intro}

Einstein-Boltzmann solvers (hereafter EBS) evolve cosmological linear perturbations for particle species (photons, baryons, cold dark matter, neutrinos, dark energy, etc.) coupled to the metric \citep{Lifshitz:1945du, Lifshitz:1963ps, Weinberg:1972kfs, Peebles:1980yev, Ma:1995ey}. EBS also deliver theoretical observables, such as CMB anisotropies and the matter power spectrum, which are necessary when fitting cosmological models to data. In modern cosmological inference, tens of millions of likelihood evaluations are routine to measure cosmological parameters, and the cumulative cost is dominated by computing these observables. EBS are therefore run massively by the worldwide cosmology community: it has been estimated that they are probably executed hundreds of billions of times every year \citep{COSMICNETI}, with substantial computational resource usage and associated electricity costs (and carbon footprint).  The most commonly used EBSs are \texttt{CAMB}\footnote{\href{https://github.com/cmbant/CAMB}{https://github.com/cmbant/CAMB}} \citep{Lewis:1999bs, camb, Howlett:2012mh} (in \texttt{FORTRAN}) and \texttt{CLASS}\footnote{\href{https://github.com/lesgourg/class_public}{https://github.com/lesgourg/class\_public}} \citep{class} (in \texttt{C}), which exhibit comparable performances. 

For minimal $\Lambda$CDM (i.e. with a cosmological constant and cold dark matter) Universes with standard settings, an EBS run can take well under a minute on a few CPU cores. Yet precise Bayesian inference, for instance, sweeps over tens of millions of models across multiple data combinations in order to provide posterior distribution for comsological parameters, turning this modest per-run time into a major bottleneck. The burden grows for extended (beyond-$\Lambda$CDM) scenarios, for instance including massive species like neutrinos \citep{Bolliet:2023sst, Jense:2024llt} or nonzero curvature \citep{Tian:2020qnm, Anselmi:2022exn, Anselmi:2022uvj}. Moreover, because of high-accuracy requirements --particularly at small angular scales-- required by future surveys, single evaluations can reach the $\sim 10$-minute regime \citep{Bolliet:2023sst, Jense:2024llt}.

Contemporary neural network-based power spectrum emulators (such as \texttt{COSMOPOWER} \citep{SpurioMancini:2021ppk} and \texttt{Capse.jl} \citep{Bonici:2023xjk}) aim to bypass repeated EBS calls, offering speedups of $\sim 10^{3\text{–}6}$. However, these gains shift the cost upstream: training typically requires $\sim 10^{5-6}$ spectra and retraining for each model variant, with the largest computational expense in non-standard cosmologies. Current (e.g. ACT \citep{Henderson:2015nzj}) and upcoming (Simons Observatory \citep{SimonsObservatory:2018koc} and Euclid \citep{euclid}) surveys   further tighten the accuracy and therefore computational demands \citep{Bolliet:2023sst, Jense:2024llt}.

In this work, we target this dual challenge—fast inference and affordable creation of training datasets for emulators—by offloading the most expensive \texttt{CAMB} modules to GPUs and making the resulting software implementation \texttt{gCAMB} publicly available~\citep{Storchi_CAMB_gpuport, Storchi_forutils_gpuport}. Beyond speed, this approach leverages comparatively underutilized GPU resources and might reduce energy consumption in certain cases, addressing also sustainability concerns alongside throughput. Notably, \texttt{gCAMB} preserves all the features of the \texttt{CAMB} code, so its adoption requires minimal changes to existing workflows. Moreover, since only the line-of-sight integration is offloaded to the GPU, the primordial and source-function modules remain essentially unchanged by the porting and can be modified as usual for testing specific theoretical models.

The paper is organized as follows. Section~\ref{sec:theory} reviews the internals of EBS, identifies GPU-amenable kernels, and outlines the scientific applications of \texttt{gCAMB}. Section~\ref{sec:comp_details} details the algorithms and implementation choices. Section~\ref{sec:results} presents profiling and validation. Section~\ref{sec:conclusion} summarizes findings and future directions.

\section{Einstein-Boltzmann solvers}\label{sec:theory}

\subsection{Key equations}\label{key_eqs}
We briefly summarize here the inner working of the EBS, with some key equations. We refer to Refs.~\citep{Ma:1995ey, Seljak:1996is} for a complete treatment in the context of cosmological perturbation theory. 

After linearizing the Einstein–Boltzmann system, Fourier perturbation modes decouple and the equations for each wavenumber $k$ can be solved independently via an ODE system in conformal time $\tau$. This yields transfer functions $\Delta_\ell^{X}(k)$; these quantities encode the relation between the anisotropies of the measurable quantity $X$ as observed from our position in space and time, and the primordial perturbations generated in the early Universe. The transfer functions are then used by the EBS to compute angular power spectra $C_\ell^{XY}$ from
\begin{equation}\label{eq:cl}
C_\ell^{XY}=4\pi\!\int\!\frac{dk}{k}\, \mathcal{P}_{\mathcal R}(k)\,\Delta_\ell^{X}(k)\,\Delta_\ell^{Y}(k)\,,    
\end{equation}
with $X,Y = \{T,E\}$, and $\mathcal{P}_{\mathcal R}(k)$ is the power spectrum of primordial perturbations. The angular power spectra are the observed quantities that can be predicted by the theory and on which cosmological likelihoods are tipically based. They describe, in harmonic space, the angular correlation between observables $X$ and $Y$. In the following, we will concentrate on CMB temperature ($T$) and $E$-mode polarization ($E$) anisotropies.

EBS compute the transfer functions in the so-called \textit{line-of-sight} formalism,
from gauge-invariant CMB source functions for temperature $S_{T}(k,\tau)$ and polarization $S_{P}(k,\tau)$ fluctuations. Assuming scalar-only perturbations sourcing only $T$ and $E$ modes (similar equations arise in the case of tensor perturbations and $B$ modes), these source functions are projected to harmonic space as \citep{Seljak:1996is}

\begin{align}
\Delta_\ell^{T}(k) &= \!\int d\tau\, S_{T}(k,\tau)\,
   \varphi_\ell\,\!\big[k(\tau_0\!-\!\tau)\big],\label{eq:transfer} \\[6pt]
\Delta_\ell^{E}(k) &= \!\int d\tau\, S_{P}(k,\tau)\,
   \epsilon_\ell\,\!\big[k(\tau_0\!-\!\tau)\big] \,. \label{eq:transferE}
\end{align}

with $\varphi_\ell,\epsilon_\ell$ kernels related to spherical Bessel functions in a spatially flat Universe (and to hyperspherical Bessel functions in a non-flat one).

\subsection{Computational bottlenecks}\label{sec:bottlenecks}

The two main bottlenecks common to EBSs (both in \texttt{CLASS} and \texttt{CAMB}) are \citep{COSMICNETI}:
\begin{enumerate}
    \item The \textit{source functions calculation}, i.e. the integration of the $k$-wise ODEs of cosmological perturbations over time $\tau$ to obtain source functions in $S_{T}(k,\tau)$ and $S_{P}(k,\tau)$;
\item  The \textit{line-of-sight integration}, i.e. the highly oscillatory, massively repeated projections $(k,\tau)\!\to\!(\ell,k)$ in Eqs.~\ref{eq:transfer}-\ref{eq:transferE} above.
\end{enumerate}
 
Which step dominates depends on context: computing CMB angular power spectra spectra for $\Lambda$CDM cosmologies up to small scales (i.e. to high multipoles), the projection stage in step 2 often wins; for extended physics (massive/interactive species, modified gravity) even basic outputs make step 1 dominant; in many $\Lambda$CDM use-cases they are comparable.

From a computational standpoint, the source function calculation in step 1 parallelizes only over the outer $k$-loop; its sequential time-stepping limits strong scaling (saturating around $\sim 10$ threads), making it a poor target for brute-force parallelization over many CPU/GPU cores. This is why neural network-based emulation of source functions—e.g., the \texttt{COSMICNET} approach \citep{COSMICNETI, COSMICNETII}—has proven effective, yielding a factor $\sim 2\!-\!3$ of end-to-end speedup for spectra while preserving accuracy. In contrast, the harmonic-transfer stage consists of vast numbers of independent integrals which is an embarrassingly parallel workload. Further algorithmic reformulations could also reduce the cost of oscillatory integrals (e.g. \citep{Assassi:2017lea, Schoneberg:2018fis}).

In what follows we focus on the \texttt{CAMB} code, targeting specifically the line-of-sight integration step with GPU porting and exploiting fine-grained parallelism at scale to accelerate it.

\section{Computational details}
\label{sec:comp_details}

In the present section, we will describe all the computational details related to the porting of the CAMB
code to the GPU. We start from a brief description of the parameter settings we used to benchmark \texttt{gCAMB} \citep{Storchi_CAMB_gpuport} against \texttt{CAMB} \citep{CAMB_fork_Storchi}.


\subsection{Cosmological and accuracy parameters for benchmarking}

To assess performance as a function of numerical accuracy, we benchmark \texttt{gCAMB} against \texttt{CAMB} under three accuracy settings -- which we refer to as \textit{Low}, \textit{Medium} and \textit{High} in the following --
while holding the cosmological parameters fixed to \textit{Planck} measured values \citep{Planck:2018vyg}. The three configurations for benchmarking are summarized in Table~\ref{tab:acc-configs}, where also the corresponding wall clock times on 32 CPU cores are reported.

In all runs we compute both scalar and tensor spectra (\texttt{get\_scalar\_cls=True}, \texttt{get\_tensor\_cls=True}). For scalars we set \texttt{l\_max\_scalar=11000} (maximum multipole for scalar $C_\ell$\footnote{Near $\ell_{\rm max}$ the accuracy of the spectra degrades; for lensed spectra one typically need some headroom above the target $\ell$ range, as recommended in the \texttt{CAMB} documentation.}) and \texttt{k\_eta\_max\_scalar=22000} (which effectively sets the wavenumber $k$
range used in line-of-sight projections for scalar perturbations); for tensors \texttt{l\_max\_tensor=1500} (maximum tensor multipole) and \texttt{k\_eta\_max\_tensor=3000} (wavenumber cutoff for tensor similarly to scalars). The three accuracy configurations use \texttt{accuracy\_boost} (i.e the global step-size/tolerance scaler\footnote{Higher values of \texttt{accuracy\_boost} tighten time steps, tolerances, and
$k$-sampling across all modules (except the ones controlled by \texttt{l\_sample\_boost} and \texttt{l\_accuracy\_boost}).}) and \texttt{l\_sample\_boost} (the output $\ell$-sampling density, i.e., number of explicitly computed multipoles\footnote{Higher values reduce interpolation error; for \texttt{l\_sample\_boost=50} \texttt{CAMB} computes all $\ell$ explicitly).}) as reported in Table~\ref{tab:acc-configs}. All other \texttt{CAMB} accuracy parameters are kept to their default value in the \texttt{params.ini} parameter file shipped with the code. We refer to the extensive \texttt{CAMB} documentation\footnote{\href{https://camb.info/}{https://camb.info/}} -- specifically we refer to the \texttt{FORTRAN} version -- for detailed explanation of all the accuracy parameters mentioned above.

\begin{table}[t]
\centering
\caption{Accuracy parameter configurations used for benchmarking and corresponding wall times for the \texttt{CAMB} code execution measured on 32 CPU cores on an Intel Xeon Platinum (PI) 8358 CPU with 2.6 GHz. See Section \ref{sec:comp_details} for details.}
\label{tab:acc-configs}
\small
\setlength{\tabcolsep}{4pt} 
\begin{tabularx}{\columnwidth}{l S[table-format=1.0] S[table-format=2.0] S[table-format=3.2]}
\hline
Setting & \texttt{accuracy\_boost} & \texttt{l\_sample\_boost} & {Wall time (32 cores)} \\
\hline
Low    & 1 &  1 &  \SI{1.56}{\second}
  \\
Medium & 3 & 20 & \SI{32.18}{\second}\\
High   & 3 & 50 & \SI{329.92}{\second} \\
\hline
\end{tabularx}
\end{table}

We increased \texttt{l\_sample\_boost} to load the GPU-accelerated line-of-sight integrator—while leaving \texttt{l\_accuracy\_boost} untouched because it would only slow down hierarchy evolution, which we do not attempt porting to GPU; this stresses the part we optimized and prepares for cases that truly need $\ell$-by-$\ell$ precision (such as primordial features, curved/anisotropic cosmologies). We note though that standard $\Lambda$CDM emulators can typically get by with the default $\sim 1.5$ sampling.

While for our benchmarks we scale accuracy also via the $\texttt{accuracy\_boost}$ parameter (following \citet{Bolliet:2023sst}), we recommend that for production runs one identifies the dominant numerical error and tune the specific sub-parameters responsible—rather than globally increasing \texttt{accuracy\_boost} — as this is typically faster and can yield higher accuracy than the uniform scaling it implies.

Regarding the need for high-accuracy settings, we note also that in practice, pushing $C_{\ell}$ to accuracy far beyond what affects post-marginalization cosmology is wasted effort and for $\ell \gg 10^3$ the dominant uncertainties are from the nonlinear lensing model—so experiment-realistic error budgets, not maximal numerical precision, should set the target.

\subsection{GPU porting of the code: \texttt{gCAMB}}

\begin{figure}[h!]
  \centering
  \begin{subfigure}[b]{0.5\textwidth}
    \includegraphics[width=\textwidth]{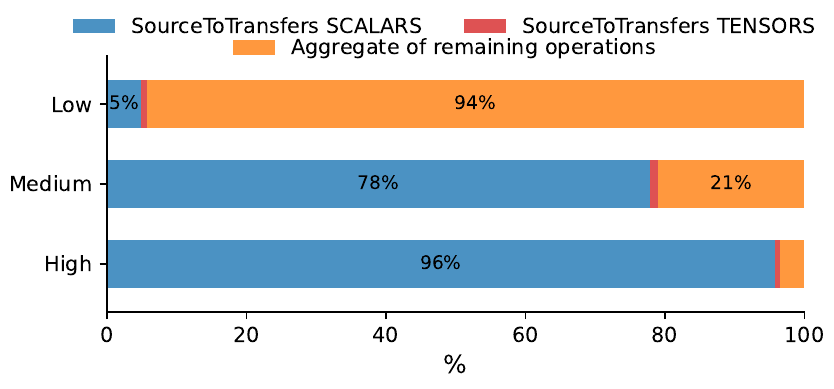}
    \caption{Serial execution: \texttt{CAMB}}
    \label{fig:element1}
  \end{subfigure}
  \hfill
   \begin{subfigure}[b]{0.5\textwidth}
    \includegraphics[width=\textwidth]{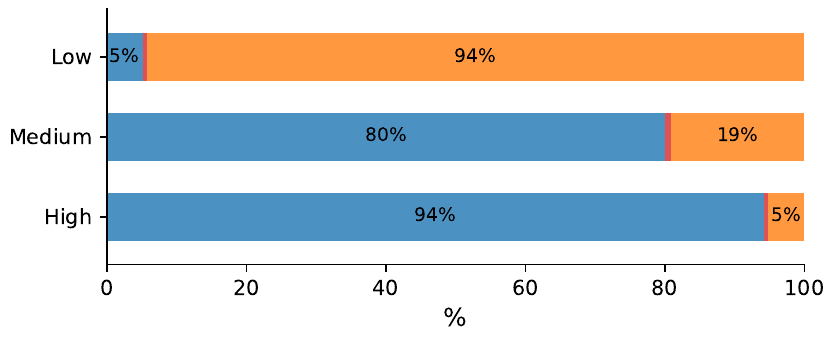}
    \caption{Multithreading (using 32 threads) execution: \texttt{CAMB}}
    \label{fig:element2}
  \end{subfigure}
  \hfill
  \begin{subfigure}[b]{0.5\textwidth}
    \includegraphics[width=\textwidth]{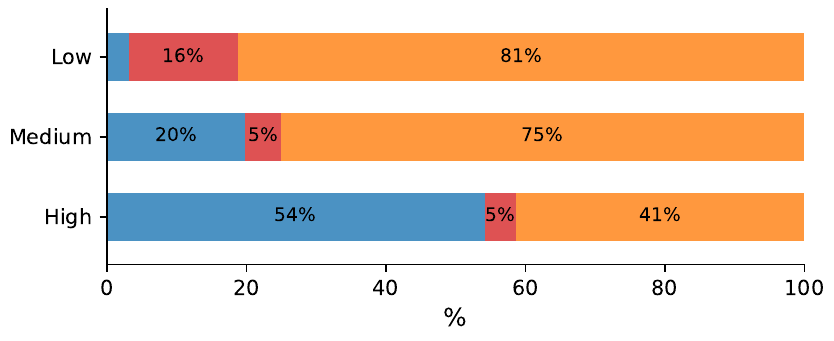}
    \caption{Hybrid GPU (NVIDIA A100) and multithreading (using 32 threads) execution: \texttt{gCAMB}}
    \label{fig:element3}
  \end{subfigure}
  \caption{Percentage of wall-time taken by the two \texttt{SourceToTransfers} tasks (one for scalar perturbations and the other for tensor ones), with 
  the remaining operations involved in the complete run reported as ``Aggregate of remaining operations''. Panel $(a)$ shows the results for the serial execution of \texttt{CAMB}, while panel $(b)$ shows \texttt{CAMB} executed on 32 OpenMP threads. Finally, panel $(c)$ shows the hybrid GPU+CPU execution of \texttt{gCAMB} on 32 threads.}
  \label{fig:percentage}
\end{figure}

As discussed in Section \ref{sec:bottlenecks}, we expect one of the two main bottlenecks in \texttt{CAMB} execution to be the line-of-sight integration of the CMB source functions, a task which we will name from now on \texttt{SourceToTransfers}. This is confirmed by profiling \texttt{CAMB} on CPUs (panel $(a)$ and $(b)$ of Figure~\ref{fig:percentage}). Details on how we profile \texttt{CAMB} are given in Section~\ref{sec:results}.

To adapt the \texttt{CAMB} code for GPU we used the OpenACC framework, a directive-based programming allowing developers to offload parts of their code to accelerators like GPUs\cite{wienke2012openacc}. We began by extracting some of the \texttt{SourceToTransfers} task related methods from their main modules such as  \texttt{CAMBmain}. We refactored these methods into standalone functions, which allowed us to precisely identify and minimize the input and output data that must be moved between the CPU and the GPU.
This strategy proved to be highly efficient, as it allowed us to create a minimal code package for the GPU, avoiding the transfer of larger, unnecessary code sections.

The final product is a streamlined code tailored specifically for the flat universe scenario. The operational flow of this ported code is summarized in the pseudo-code reported in Algorithm \ref{alg:minimal_source_transfers}, where we outline the primary \texttt{SourceToTransfers} task for the accelerator not encompassing the complete calling sequence or all operations, but rather aiming to provide a concise overview of the ported procedure. Indeed, the overall computation is handled by the loop over the number of points that is implemented in the \texttt{TimeSourcesToCl} subroutine within the \texttt{CAMBmain} module. Finally, it is the main routine \texttt{SourceToTransfers}, within the main loop, that orchestrates the entire process. This routine begins by calling \texttt{IntegrationVars\_init} to set up the necessary integration variables and \texttt{InterpolateSources} to prepare the source data. The core of the calculation is performed by the \texttt{DoSourceIntegration} function. Within this function, the majority of the computational effort is spent in the \texttt{DoFlatIntegration} subroutine.  
The whole procedure can be summarized as follow: 

\begin{itemize}
    \item Data Transfer to GPU: The process starts by allocating memory on the GPU and copying the necessary input data from the main system's memory (CPU) to the GPU's memory. This is represented by the ``begin accelerator data region'' block in the pseudo-code~\ref{alg:minimal_source_transfers}.
    \item Coarse-Grained Parallelism (Gang): The outer loop runs over the index $q_{ix}=1,\dots,N$, each corresponding to a cosmological perturbation mode. In flat geometry, this is the comoving wavenumber $k$. This loop is parallelized at a high level (gang). Each iteration, which involves a call to the \texttt{SourceToTransfers} routine, is treated as an independent task assigned to a group of processing units on the GPU.
    \item Fine-Grained Parallelism (Vector): Inside the \texttt{SourceToTransfers} routine, there are several loops, of which the most demanding ones are the two nested inner loops reported in the pseudo-code. The first one is a vectorized loop over $j=1,\dots,L_{\max}$ corresponding to cycling over the multipoles $\ell$. For each $\ell$ and source $X$ (e.g., temperature, $E$/$B$ polarization, lensing), the routine builds the transfer functions $\Delta^{X}_\ell(q)$. Nested within the $\ell$-loop is a vectorized loop over $n=n_{\min},\dots,n_{\max}$ that cycles through discretized line-of-sight sampling points (conformal times $\eta_n$ or equivalently comoving distances $\chi_n$). This loop accumulates the line-of-sight integral in Eqs.~\ref{eq:transfer}, which involves calculation of the spherical Bessel-function. All of these loops are parallelized at a much finer level (vector). This means that the iterations within these loops are executed simultaneously by individual processing threads within a gang, which is highly efficient for array-based calculations such as the Bessel integration mentioned above. 
    \item Data Transfer from GPU: Once all the parallel computations are finished, the resulting transfer functions—the 3D array $\{\Delta^{X}_\ell(q)\}$ over sources $X$, multipoles $\ell$, and modes $q$— are copied from the GPU's memory back to the CPU's memory, and the memory on the GPU is freed up. This is marked by the ``end accelerator data region'' command in the pseudo-code reported in Algorithm ~\ref{alg:minimal_source_transfers}.  
\end{itemize}

\begin{algorithm}
\caption{The following pseudocode outlines the primary \texttt{SourceToTransfers} task ported to the GPU. This representation does not encompass the complete calling sequence or all operations, but rather aims to provide a concise overview of the ported procedure. See the text for a more detailed description of the full calling sequence.}
\label{alg:minimal_source_transfers}
\begin{algorithmic}[1]

\State \textbf{begin} accelerator data region \Comment{Copy data to GPU}
\For{each $q_{ix}$ from 1 to N} \textbf{in parallel (gang)}
    \State \uline{\Call{SourceToTransfers}{$q_{ix}$, ...}}
        \Statex \Comment{\textit{\qquad Inside SourceToTransfers (executed on GPU)}}
        \For{each $j$ from 1 to $L_{max}$} \textbf{in parallel (vector)}
            \State ... \Comment{Initialize local sums}
            \For{each $n$ from $n_{min}$ to $n_{max}$} \textbf{in parallel (vector)}
                \State \Comment{Perform calculations (e.g., Bessel integration)}
                \State ...
            \EndFor
        \EndFor
\EndFor
\State \textbf{end} accelerator data region \Comment{Copy results from GPU}
\end{algorithmic}
\end{algorithm}

\section{Results and discussion}\label{sec:results}

This section provides a detailed analysis of the numerical and timing results obtained from the \texttt{gCAMB} code. Our investigation focuses on evaluating the efficiency and accuracy of the \texttt{gCAMB} implementation, particularly in comparison to other existing versions or methodologies.

\subsection{Computational environment and build configuration}

All tests were performed with the Tier-0 EuroHPC supercomputer Leonardo at CINECA \citep{turisini2023leonardopaneuropeanpreexascalesupercomputer}
running an Intel Xeon Platinum (PI) 8358 CPU with 2.6 GHz (total 32 cores), in the case of the accelerated CPU/GPU version we use a single NVIDIA A100 GPU.
When compiling the \texttt{gCAMB} code, we use the NVIDIA HPC compiler (nvfortran version 24.5-1) for CPU and CPU/GPU. Instead, we used the GNU Fortran compiler, version 8.5.0, when compiling the serial and the OpenMP (CPU only) versions of the code. This is due to a known issue of the NVIDIA \texttt{FORTRAN} compiler, detailed in Section~\ref{sec:timing}. 

\subsection{Benchmark configurations and measured quantities}

Table~\ref{tab:timing} provides a comprehensive overview of the performance characteristics of various code versions, focusing on the total execution time (walltime) and the specific time consumed by the crucial \texttt{SourceToTransfers} task. Given that both scalar and tensorial modes are processed, the \texttt{SourceToTransfers} procedure is invoked twice; consequently, both the individual timings for each execution and their cumulative duration are reported.
The versions considered in this analysis include: ``Serial (CPU)'', ``OpenMP (32 thr)'', ``\texttt{gCAMB} (GPU+CPU, 32 thr)'' and ``\texttt{gCAMB} (GPU+CPU, 32 thr, est.)''. The ``Serial (CPU)'' represents the baseline, standard implementation of the \texttt{CAMB} code, executed sequentially without parallelization. It serves as a fundamental point of comparison for evaluating the benefits of parallel computing. the ``OpenMP (32 thr)'' version leverages the Open Multi-Processing (OpenMP) application programming interface for parallel programming, specifically targeting CPU-only execution. We utilize 32 threads to distribute computational tasks across multiple cores. The ``\texttt{gCAMB} (GPU+CPU, 32 thr)'' represents a hybrid approach, integrating both GPU and CPU resources for computation. Like the OpenMP version, it also employs 32 threads, demonstrating an effort to accelerate performance by offloading the \texttt{SourceToTransfers} task to the GPU while retaining CPU involvement for other tasks.

\subsection{Estimated fully OpenMP-enabled timing}\label{sec:timing}

Additionally, Table~\ref{tab:timing} includes a specific entry, ``\texttt{gCAMB} (GPU+CPU, 32 thr, est.)'', which presents an estimated walltime for a hypothetical, fully OpenMP-ported version of \texttt{gCAMB}. This estimation is a necessary inclusion due to current technical limitations encountered with the NVIDIA \texttt{FORTRAN} compiler. Specifically, the compiler exhibits restrictions regarding the proper setup of polymorphic types within certain OpenMP regions. These limitations necessitated the temporary removal of some OpenMP sections during the \texttt{gCAMB} compilation process, preventing a direct, fully OpenMP-enabled measurement. The estimated value thus provides valuable insight into the potential additional performance gains achievable once these compiler constraints are overcome and a complete OpenMP integration is feasible.

Estimation is achieved by first calculating the total walltime recorded for the OpenMP execution. From this total walltime, the duration attributed to \texttt{SourceToTransfers} operations is subtracted. Subsequently, the \texttt{SourceToTransfers} computation time derived specifically from the \texttt{gCAMB} code is instead added. This method aims to crudely estimate the walltime for a hypothetical, fully OpenMP-ported version of \texttt{gCAMB}. In general, the NVIDIA compilation issue is projected to affect less than 10\% of the high- and low-accuracy scenarios. However, its influence is more significant in the medium accuracy scenario, leading to a performance reduction of up to 38\%.

\begin{table*}[t]
\centering
\caption{Wall-clock times (seconds) for \texttt{CAMB}/\texttt{gCAMB} under the three accuracy settings (low, medium and high) described in Section \ref{sec:comp_details}. For each setting, we report the execution times of \texttt{CAMB} on a single CPU core (``Serial (CPU)'') and on 32  OpenMP CPU threads (``OpenMP (32 thr)''), hybrid GPU and CPU execution of \texttt{gCAMB} (``\texttt{gCAMB} (GPU+CPU, 32 thr)'') and the estimated times for hybrid execution of \texttt{gCAMB} expected for the full OpenMP port (i.e. ``\texttt{gCAMB} (GPU+CPU, 32 thr, est.)''). For each execution method, we report the runtimes of the \texttt{SourceToTransfers} routine for scalars and tensors separately and together (``Subtotal''). Finally, the last column (``All'') shows the runtimes for end-to-end execution for each method. For further details see Section \ref{sec:results}. 
All runs were performed on the Leonardo supercomputer at CINECA.}
\label{tab:timing}
\begin{threeparttable}

\normalsize
\setlength{\tabcolsep}{3.5pt}   
\renewcommand{\arraystretch}{0.95}
\sisetup{table-number-alignment = center, table-figures-integer = 4, table-figures-decimal = 2}

\makebox[\textwidth][c]{%
  \hspace*{0.06\textwidth}
  \begin{tabularx}{0.80\textwidth}{@{} 
      l
      >{\raggedright\arraybackslash}p{0.34\textwidth} 
      @{\hskip 2pt} 
      S[table-format=4.2]
      S[table-format=5.2]
      S[table-format=5.2]
      S[table-format=5.2]
    @{}}
  \toprule
  \multirow{2}{*}{Accuracy} & \multirow{2}{*}{Run method} & \multicolumn{3}{c}{\texttt{SourceToTransfers} [s]} & {All [s]} \\
  \cmidrule{3-5}\cmidrule{6-6}
   &  & {Scalars} & {Tensors} & {Subtotal} & {Total} \\
  \midrule
  \multirow{4}{*}{Low}
    & Serial (CPU)                    & 1.32   & 0.22 & 1.54   & 26.79 \\
    & OpenMP (32 thr)                 & 0.08   & 0.01 & 0.09   & 1.56  \\
    & \texttt{gCAMB} (GPU+CPU, 32 thr)& 0.06   & 0.29 & 0.35   & 1.86  \\
    & \texttt{gCAMB} (GPU+CPU, 32 thr, est.) & 0.06 & 0.29 & 0.35   & 1.83  \\
  \addlinespace[4pt]
  \midrule
  \multirow{4}{*}{Medium}
    & Serial (CPU)                    & 457.98 & 6.94 & 464.92 & 588.19 \\
    & OpenMP (32 thr)                 & 25.76  & 0.28 & 26.04  & 32.18  \\
    & \texttt{gCAMB} (GPU+CPU, 32 thr)& 3.33   & 0.85 & 4.18   & 16.78  \\
    & \texttt{gCAMB} (GPU+CPU, 32 thr, est.) & 3.33 & 0.85 & 4.18   & 10.33  \\
  \addlinespace[4pt]
  \midrule
  \multirow{4}{*}{High}
    & Serial (CPU)                    & 5983.50 & 43.92 & 6027.42 & 6246.63 \\
    & OpenMP (32 thr)                 & 310.67  & 1.92  & 312.59  & 329.92  \\
    & \texttt{gCAMB} (GPU+CPU, 32 thr)& 29.24   & 2.46  & 31.70   & 54.01   \\
    & \texttt{gCAMB} (GPU+CPU, 32 thr, est.) & 29.24 & 2.46 & 31.70   & 49.04   \\
  \bottomrule
  \end{tabularx}%
} 
\end{threeparttable}
\end{table*}

\subsection{Performance results and speedups}

\begin{figure}[h!]
  \centering
  \begin{subfigure}[b]{0.5\textwidth}
    \includegraphics[width=\textwidth]{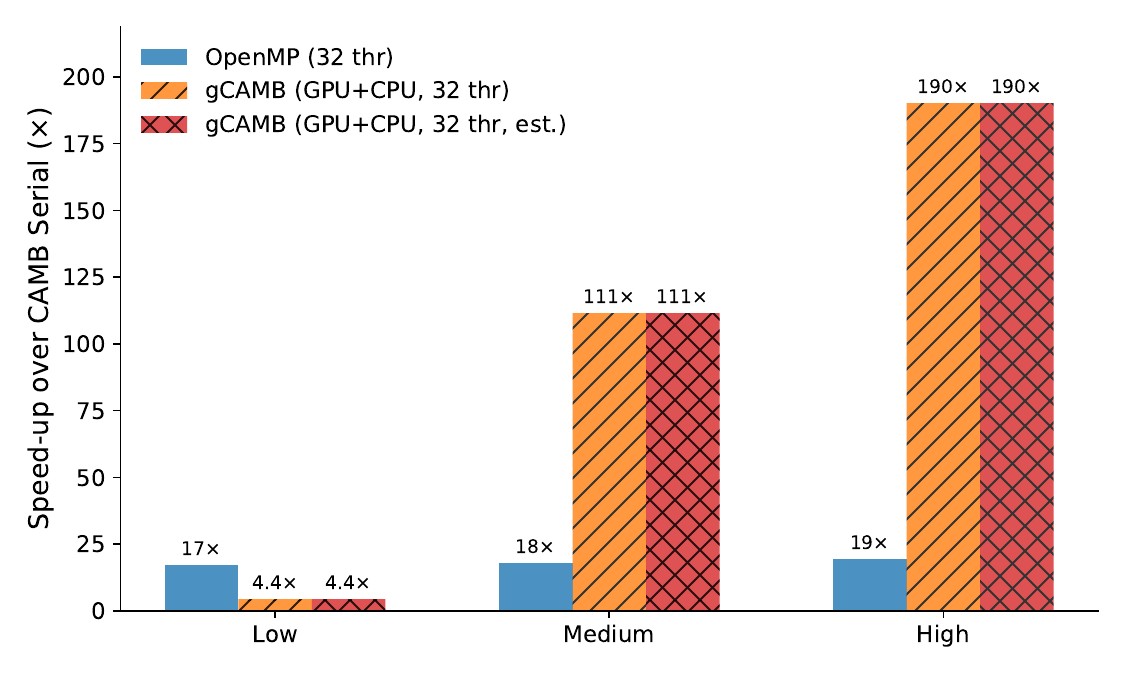}
    \caption{\texttt{SourceToTransfers} Speed-up}
    \label{fig:element1}
  \end{subfigure}
  \hfill
  \begin{subfigure}[b]{0.5\textwidth}
    \includegraphics[width=\textwidth]{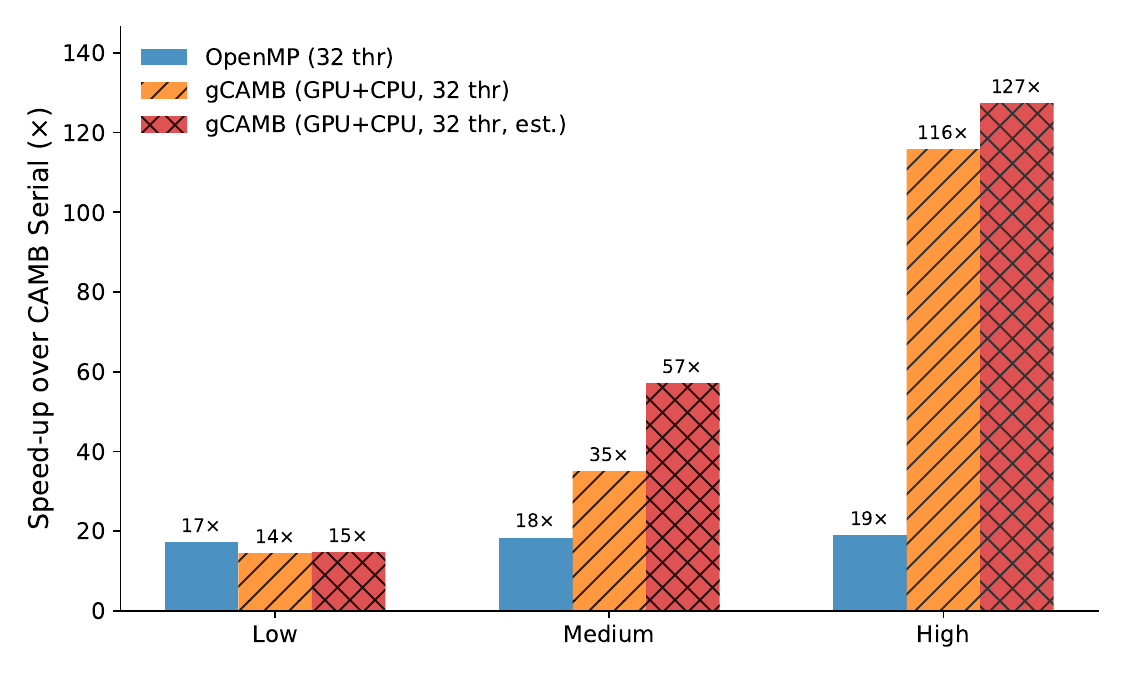}
    \caption{Overall Speed-up}
    \label{fig:element2}
  \end{subfigure}
  \caption{Speed-up factors relative to the serial \texttt{CAMB} execution on a single CPU core for both the \texttt{SourceToTransfers} task (upper panel) and the end-to-end execution (``Overall''). We show the speed-up factors for the low, medium and high accuracy settings and for \texttt{CAMB} parallel execution on 32 CPU cores (``OpenMP 32 Threads''), hybrid GPU and CPU execution of \texttt{gCAMB} (``\texttt{gCAMB} 32 Threads)'') and the estimated times for hybrid execution of \texttt{gCAMB} expected for the full OpenMP port (i.e. ``\texttt{gCAMB} 32 Threads (Est.)'').}
  \label{fig:speedup}
\end{figure}

Based on the timing data in Table \ref{tab:timing} and the speed-ups shown in Figure \ref{fig:speedup}, the primary advantage of using \texttt{gCAMB} is the dramatic acceleration of the \texttt{SourceToTransfers} task, which is the main computational bottleneck in the standard \texttt{CAMB} code. This targeted optimization leads to a very significant reduction in the overall time required to generate cosmological power spectra, especially for 
the high parameters configuration. Indeed, as shown in Figure \ref{tab:timing}, the speed-up for the \texttt{SourceToTransfer} task alone is enormous. For the high accuracy setting, \texttt{gCAMB} achieves a speed-up of nearly 10x compared to the multi-threaded CPU version (OpenMP 32 Threads). 
Because \texttt{SourceToTransfers} dominates the total runtime (especially in medium and high accuracy scenarios, see also Figure \ref{fig:percentage}), accelerating it provides a substantial overall performance boost. Figure \ref{fig:speedup} shows that for the high accuracy setting, the overall speed-up is over 100x. This transforms a calculation that takes over 5 minutes on 32 CPU cores (329.92s) into one that completes in under a minute (54.01s) on the GPU.

\subsection{GPU warm-up behavior and the low-accuracy regime}

It is clear that the advantage of \texttt{gCAMB} becomes more pronounced as the computational demand increases. Indeed, in the low accuracy scenario, the overhead of GPU warm-up time makes the benefit less obvious. It is a common and expected phenomenon for the first execution of a GPU-accelerated kernel to  be significantly slower than subsequent runs, even when performing the exact same task.  This initial delay is often referred to as a ``warm-up'' period and is caused by a combination of one-time setup and initialization overheads. Reason why the speed-up of the first \texttt{SourceToTransfers} task is generally always the
worst respect to the second \texttt{SourceToTransfers} task execution. Clearly while in the medium and high accuracy scenarios the improvement in terms of speed-up in the tensors
GPU execution is able to fully recover this warm-up time, this is not the case for the low accuracy scenario. 

\subsection{Runtime composition and remaining acceleration opportunities}

It should be noted that the computational load of \texttt{SourceToTransfers} constitutes, excluding the case of the low-accuracy scenario, a significant portion of the overall calculation, see Figure \ref{fig:percentage}. Furthermore, it is evident that this percentage shifts considerably when the primary computational load is offloaded to the GPU, with the exception of the low accuracy scenario, which is influenced by the aforementioned ``warm-up'' period. Overall, this demonstrates that there is potential for further improvement by porting additional sections of the code to the GPU.

\subsection{Reduction in power consumption}
Before we start reporting the numerical results, it is also notable that we conducted a preliminary estimation of the power consumption of the \texttt{CAMB} code compared to the \texttt{gCAMB} one. Software measurements were performed on a machine equipped with an NVIDIA RTX 6000 Ada Generation and a dual AMD EPYC 9224 24-Core Processor, so that we could run performance analysis tools (as detailed below) with root permissions, which is not possible on the Leonardo computer cluster.  Using 32 threads, for both \texttt{CAMB} and \texttt{gCAMB} execution, a significant decrease in power consumption was observed in the high accuracy scenario. In this scenario, the CPU-only code (\texttt{CAMB}) consumed 61283.24 Joules, while the \texttt{gCAMB} code, accounting for both CPU and GPU consumption, consumed 33042.42 Joules, representing an approximate halving of power consumption. All measurements were conducted using the \texttt{perf} tool \citep{deMelo:2010:NPT} for CPU energy consumption and \texttt{nvidia-smi} one \citep{NVIDIA:smi} for GPU consumption (the energy consumption metric was not available for this GPU via the NVIDIA Nsight Compute tool
\citep{NVIDIA:NsightCompute}). 


\subsection{Numerical validation of power spectra}
We quantify  the numerical agreement between the spectra obtain from \texttt{CAMB} running on 32 OpenMP CPU threads and \texttt{gCAMB} hybrid GPU+CPU execution 32 OpenMP threads by the fractional difference per multipole,
$$
\frac{\Delta C_\ell^X}{C_\ell^X}\equiv\frac{C_\ell^{X,\mathrm{GPU}}-C_\ell^{X,\mathrm{CPU}}}{C_\ell^{X,\mathrm{CPU}}},
\quad X\in\{\mathrm{TT,TE,EE,BB}\}.
$$

\begin{figure*}[h!]
    \centering
    \includegraphics[width=0.9\linewidth]{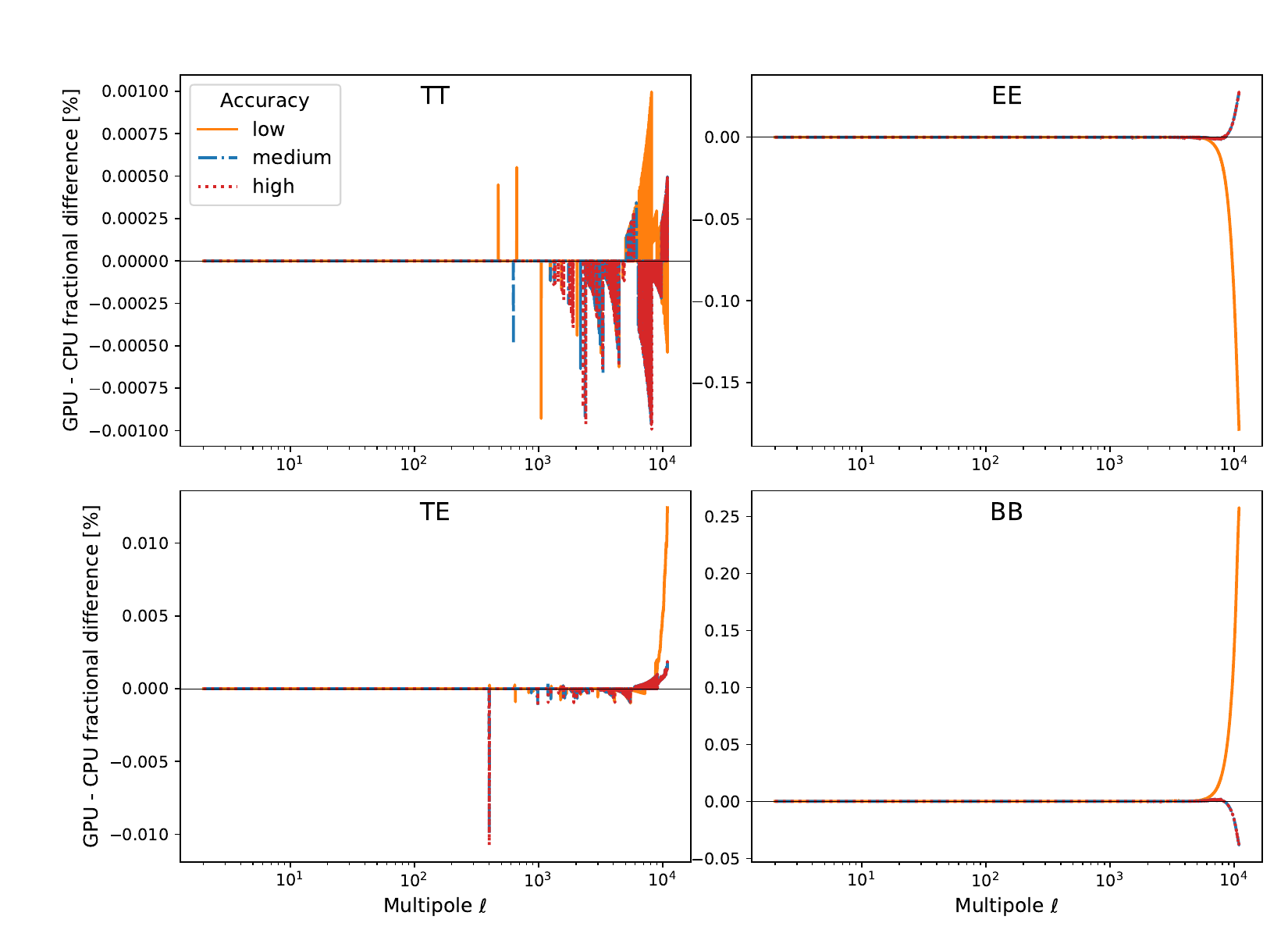}
  \caption{Fractional difference (in percentage) in low, medium and high accuracy spectra between \texttt{gCAMB} running on GPU and \texttt{CAMB} on CPUs.}
  \label{fig:difference_spectra}
\end{figure*}

The fractional differences (one per spectrum, considering the usual low, medium and high accuracy settings) are shown in Fig. \ref{fig:difference_spectra}.

Additionally, we compare the difference between GPU- and CPU-computed spectra to the cosmic variance (assuming an ideal noiseless full-sky observation) $\sigma_{\rm CV}^{XY}(\ell)$ at each multipole $\ell$. If $|\Delta C_\ell^{XY}|\ll \sigma_{\rm CV}^{XY}(\ell)$ we considered this test passed for \texttt{gCAMB}. The cosmic variance for auto (TT, EE, BB) and cross (TE) spectra\footnote{We do not consider $TB$ and $EB$ here because they are expected to vanish in standard scenarios.} is given by

\begin{align}
 \sigma_{\rm CV}^{TT,EE,BB}(\ell)&=\sqrt{\frac{2}{2\ell+1}}\;\big|C_\ell^{TT,EE,BB}\big|\,, \\[6pt]
 \sigma_{\rm CV}^{TE}(\ell)&=\sqrt{\frac{(C_\ell^{TE})^2+C_\ell^{TT}C_\ell^{EE}}{2\ell+1}}.
\end{align}

\begin{figure*}[h!]
    \centering
    \includegraphics[width=0.9\linewidth]{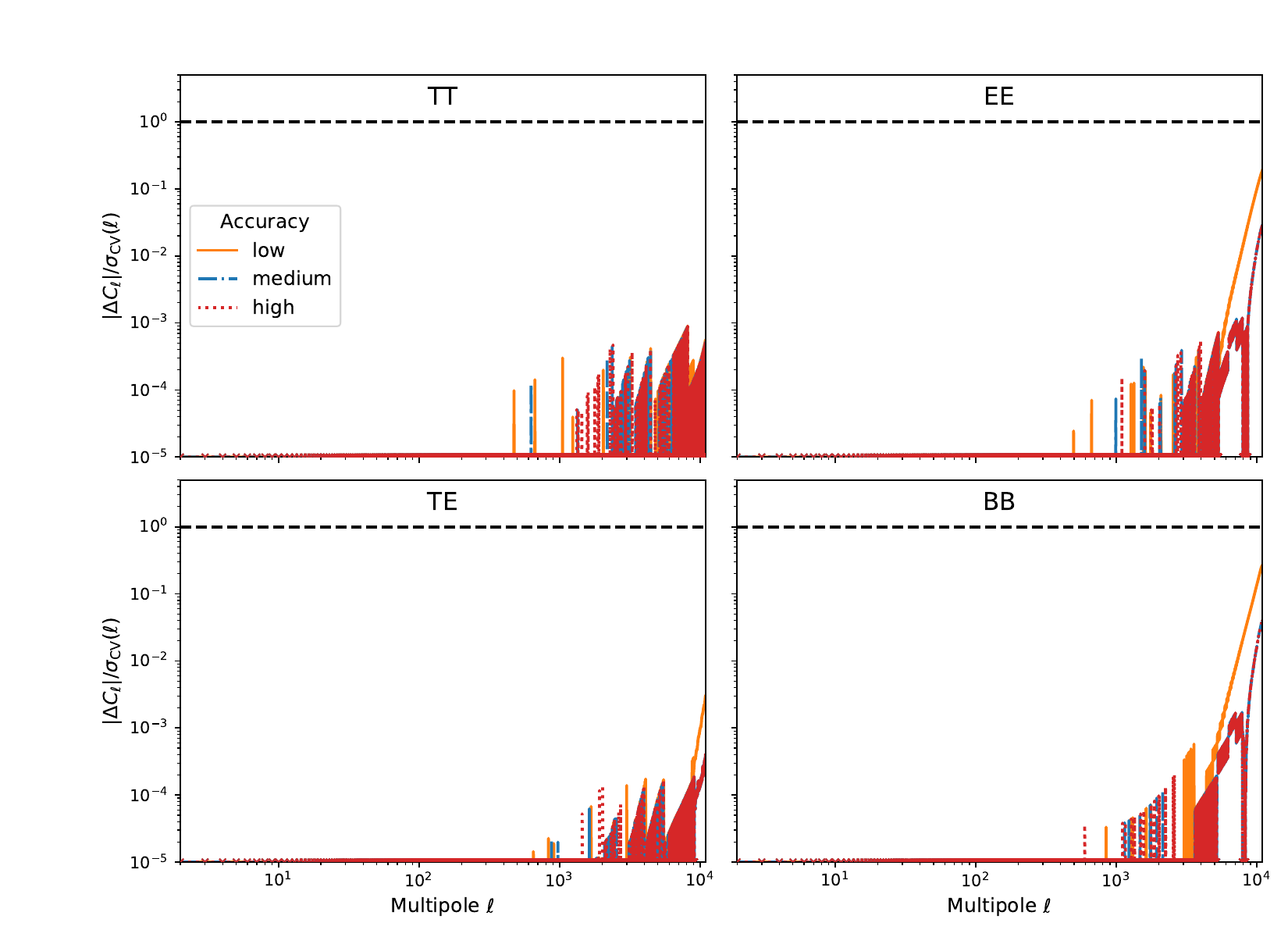}
  \caption{
   Ratio $|\Delta C_\ell|/\sigma_{\rm CV}(\ell)$ of the absolute value of the difference in low, medium and high accuracy spectra between \texttt{gCAMB} running on GPU and \texttt{CAMB} on CPUs and the cosmic variance computed on full-sky at each multipole. The dashed black horizontal line at 1 marks the strict per–multipole threshold given by cosmic variance. The numerical difference between \texttt{gCAMB} and \texttt{CAMB} is always below the cosmic variance at least by a factor 5.}
  \label{fig:cosmic_var}
\end{figure*}

The $|\Delta C_\ell|/\sigma_{\rm CV}(\ell)$ for all spectra and the different accuracy settings is shown in Fig.~\ref{fig:cosmic_var}, with the horizontal line at 1 marking the per–multipole cosmic-variance threshold.

Across all spectra and all multipoles we find $|\Delta C_\ell|/\sigma_{\rm CV}(\ell)<1$. The worst case occurs at the highest multipoles in $EE$ and $BB$, where the numerical differences still remain at least a factor of $\sim5$ below full-sky cosmic variance; in all other regimes the differences are several orders of magnitude smaller. Although Fig.~\ref{fig:cosmic_var} shows $\Delta C_{\ell}$ well below cosmic variance, this is only a sanity check—not a guarantee of unbiased inference—because coherent same-sign residuals (even at the $\sim 0.1\%$ level for cosmic variace-limited data to $\ell$) can bias parameters, so the proper validation is a posterior bias test rather than per-$\ell$ cosmic variance-limit comparisons. Real measurements further degrade per–mode precision --due for instance to finite $f_{\rm sky}$, beams, anisotropic noise, foregrounds, and systematics -- experimental uncertainties per-multipole will be substantially larger in current and future experiments, making any residual numerical mismatch even less consequential.

\section{Conclusions and future prospects}\label{sec:conclusion}

The extensive profiling we performed indicates that \texttt{gCAMB} achieves a speedup of a factor $\sim 10$ on an NVIDIA A100 compared with standard \texttt{CAMB} on 32 CPU cores. This substantially reduces the compute time required to generate typical emulator training sets of $10^{5\text{--}6}$ spectra. Validation tests also show that numerical differences introduced by the GPU port are fully negligible and remain well below the cosmic variance limit on full-sky at each multipole. This makes the adoption of \texttt{gCAMB} completely safe from numerical errors for next-generation surveys even at very high multipoles (small scales). Crucially, \texttt{gCAMB} is a feature-complete, drop-in replacement for \texttt{CAMB}, so migrating existing workflows typically requires very few changes. Moreover, the GPU port isolates only the line-of-sight projection, leaving the primordial and source-function modules unchanged—and therefore editable --as commonly done for \texttt{CAMB}-- for testing specific theoretical models.

We therefore expect \texttt{gCAMB} to be useful to the community for producing large, high-accuracy spectral datasets more quickly, with reduced CPU hours consumption and improved sustainability.
Indeed, a notable reduction in power consumption was observed in the high accuracy scenario, where the power consumption estimated by the \texttt{gCAMB} code is approximately half that of the CPU-only code. We note that further energy benchmarking on several different CPU and GPU models is needed, as this could depend on the specific hardware used here.
 As immediate next steps, porting also the \texttt{DoNonFlatIntegration} routine to the GPU should further increase throughput and make, in principle, systematic explorations of curved universes feasible and significantly cheaper. In addition, integrating \texttt{gCAMB} with sampling algorithms (e.g., MCMC) would enable direct evaluation of theoretical spectra without emulators when desired.

Among other possibilities, we envisage the following potential applications of \texttt{gCAMB}.
Firstly, as anticipated, \texttt{gCAMB} is indeed well suited to producing massive datasets of extremely high-accuracy spectra to train robust power-spectrum emulators \citep{SpurioMancini:2021ppk, Jense:2024llt}. This is advantageous for standard $\Lambda$CDM and even more valuable for costlier non-flat cosmologies and extended models.

Models with sharp or resonant primordial features (e.g., axion–monodromy \citep{Zeng:2018ufm} or EFT “standard clock’’ signals \citep{Aslanyan:2014mqa, Calderon:2025xod}) imprint rapid oscillations in $\Delta_\ell^X(k)$ and $C_\ell$ \citep[see also][]{Braglia:2021ckn}. Computing every multipole $\ell$ without spline interpolation—and evaluating transfer kernels on dense $k$ grids—improves fidelity in these scenarios. High-throughput GPU line-of-sight projections in \texttt{gCAMB} make such dense sampling practical for targeted searches in CMB and LSS.

Non-flat models replace spherical Bessel kernels with hyperspherical Bessel functions and are substantially more expensive \citep{Lesgourgues:2013bra, Kosowsky:1998nc, Anselmi:2022exn, Anselmi:2022uvj, Tian:2020qnm}. Efficient evaluation of curved-space transfer functions and line-of-sight projections is a known bottleneck; \texttt{gCAMB} can accelerate these steps, enabling systematic scans over $\Omega_K \neq 0$.

\texttt{gCAMB} could in principle be useful also for exploring anisotropic/topological and other beyond-FLRW signatures \citep{COMPACT:2022gbl}.
Statistical anisotropy or nontrivial topology couples modes and often demands fine sampling in both $k$ and $\ell$. Avoiding interpolation artifacts is particularly valuable when testing such models against large-angle CMB anomalies or anisotropic clustering statistics.

\section*{Acknowledgements}
We thank Stefano Anselmi,  Antony Lewis and Glenn Starkman for useful comments and discussion.
We acknowledge the use of computing facilities provided by the INFN theory group (I.S. InDark) at CINECA. 

This work has been supported by the Fondazione ICSC, Spoke 3 Astrophysics and Cosmos Observations. National Recovery and Resilience Plan (Piano Nazionale di Ripresa e Resilienza, PNRR) Project ID CN\_00000013 ``Italian Research Center on High-Performance Computing, Big Data and Quantum Computing''  funded by MUR Missione 4 Componente 2 Investimento 1.4: Potenziamento strutture di ricerca e creazione di ``campioni nazionali di R\&S (M4C2-19 )'' - Next Generation EU (NGEU). 

M.L. acknowledge the financial support from the INFN InDark initiative and from the COSMOS network through the ASI (Italian Space Agency) Grants 2016-24-H.0 and 2016-24-H.1-2018. M.L. is funded by the European Union (ERC, RELiCS, project number 101116027). 

L. S. and N. A. acknowledge funding from “PaGUSci - Parallelization and GPU Porting of Scientific Codes”  CUP: C53C22000350006 within the Cascading Call issued by Fondazione  ICSC , Spoke 3 Astrophysics and Cosmos Observations.

\appendix



\bibliographystyle{elsarticle-harv} 
\bibliography{example}






\end{document}